# Measuring Method of a Surface Property inside the Pore: Application of Kelvin's equation

Ken-ichi Amano[a]

[a]*Department of Energy and Hydrocarbon Chemistry, Graduate School of Engineering, Kyoto University, Kyoto 615-8510, Japan.*

E-mail: amano.kenichi.8s@kyoto-u.ac.jp

**ABSTRACT**

Surface analyses inside the nanopore, micropore, and a very narrow pipe are important topics for development of the chemical engineering. Here, we propose a measuring method which evaluates the surface coverage of the chemically modified pore surface and the corrosion rate of the inner surface of the narrow pipe, etc. The method uses Kelvin's equation that expresses saturated vapor pressure of a liquid in the pore (pipe). The surface coverage and the corrosion rate are calculated by measuring saturated vapor pressure of the liquid in the pore and the pipe, respectively. In this letter, we explain the concept of the method briefly.

## MAIN TEXT

Surface analyses inside the nanopore, micropore, and very narrow pipe are important topics for development of the chemical engineering (e.g., porous material, micro flow channel, battery material, dialysis membrane, etc). Here, we propose a measuring method which evaluates the surface coverage of the chemical modification, adsorption rate of a chemical substance, or corrosion rate of the inner surface of the narrow pipe. The method uses Kelvin's equation [1-7] that expresses saturated vapor pressure (SVP) of a liquid in the pore as follows:

$$P = P_0 \exp\left(-\frac{4V_l \gamma_{gl} \cos\theta}{DRT}\right), \tag{1}$$

where $P$ is SVP of the liquid in the pore, $P_0$ SVP of the bulk liquid, $V_l$ molar volume of the liquid, $\gamma_{gl}$ the liquid/gas surface tension, $\theta$ the contact angle of the liquid on the plane surface which is composed of the inner surface of the pore, $D$ diameter of the pore (pipe), $R$ gas constant, $T$ temperature. Recently, it is confirmed by Mima *et al.* [8] that Kelvin's equation is accurate even if the pore (pipe) diameter is several nanometers. Thus, Kelvin's equation can be used in the cases of nanopore and micropore. From Eq. (1), $\cos\theta$ is expressed as

$$\cos\theta = \frac{DRT}{4V_l \gamma_{gl}} \ln\left(\frac{P_0}{P}\right). \tag{2}$$

By inserting the liquid into the pore, $P$ is experimentally obtained. Then, substituting $P$ into Eq. (2), $\cos\theta$ and the contact angle $\theta$ are calculated. To calculate the surface coverage, adsorption rate, or corrosion rate, the contact angle obtained here is employed. The calculation process is explained in the next section. Hereafter, the surface coverage, adsorption rate, and corrosion rate are, as a whole, called inner-surface property (ISP).

ISP of the pore surface being $\sigma$ is estimated by preparing plane surfaces having several values of ISP. ISP of those plane surfaces $\sigma_P$ are measured by an arbitrary experimental method (e.g., Scanning Probe Microscope, X-ray photoelectron Spectroscopy, Infrared Spectroscopy, etc). Furthermore, the contact angles of those plane surfaces $\theta_P$ are also measured beforehand for estimation of $\sigma$. Then, plotting the results $\sigma_P$ and $\theta_P$, the experimental curve ($\sigma_P$ vs $\theta_P$) is obtained (Fig. 1). Finally, referring the experimental curve ($\sigma_P$ vs $\theta_P$) and using $\theta$, ISP inside the pore $\sigma$ is evaluated. This is the brief explanation of the method.

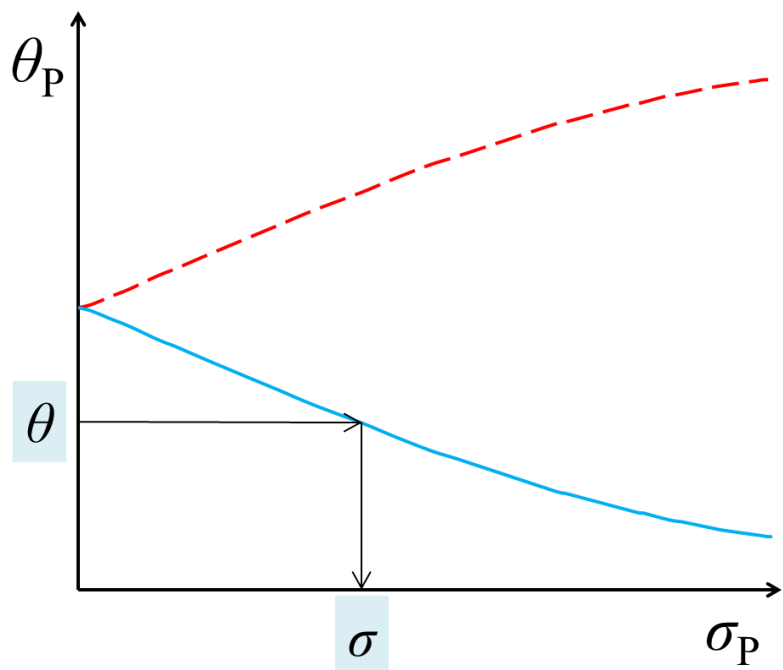

Fig. 1: *Schematic* of the $\sigma_P$ vs $\theta_P$. If the modified molecule is solvophilic, the contact angle $\theta_P$ decreases as the surface coverage $\sigma_P$ increases (Blue solid line). If the modified molecule is solvophobic, the contact angle $\theta_P$ increases as the surface coverage $\sigma_P$ increases (Red broken line). Using $\theta$ obtained from SVP of the solvent in the pore (employ Eq. (2)), the surface coverage inside the pore $\sigma$ is estimated.

When environment inside the pore (pipe) is super hydrophobic, it is difficult for water to enter the pore (pipe), and thus SVP cannot be measured. Moreover, when the pore (pipe) environment is super hydrophilic, it is difficult to measure SVP due to almost zero vapor pressure. Theoretically, however, Kelvin's equation is applicable to various kinds of liquids. Hence, it is possible to assess the inner surfaces of the super hydrophobic and hydrophilic pores (pipes) by using *a liquid other than water* (e.g., organic solvent).

In summary, we have proposed a method for evaluating the inner surface property of the nanopore, micropore, and very narrow pipe. It is expected that the surface coverage, adsorption rate, and corrosion rate can be obtained by following the method above. The method proposed here is just a theoretical one, and it has not been practically tested yet. Thus, in the future, we should perform the experiment.

**ACKNOWLEDGEMENTS**

We appreciate discussions with T. Sakka, N. Naoya, K. Fukami, and R. Koda.

**REFERENCES**

[1] V. K. La Mer and R. Gruen, Trans. Faraday. Soc. **48**, 410 (1952).

[2] L. M. Skinner and J. R. Sambles, Aero. Sci. **3**, 199 (1972).

[3] L. R. Fisher, J. Colloid Interface Sci. **80**, 528 (1981).

[4] G. Mason, J. Colloid Interface Sci. **80**, 36 (1982).

[5] J. P. Powles, J. Phys. A: Math. Gen. **18**, 1551 (1985).

[6] W. Wu and G. H. Nancollas, J. Sol. Chem. **27**, 521 (1998).

[7] J. S. Rowlinson and B. Widom, *Molecular theory of capillary*, Courier Dover Publications (2002).

[8] T. Mima *et al.*, The 28th Annula Meeting of the Molecular Simulation Society of Japan, 147P (page 184), Nov. 12-14 (2014) Sendai, Japan.